\documentclass[12pt]{iopart} 
\usepackage{amsfonts,amssymb}
\usepackage{epsfig}
\usepackage{iopams,cite}
\usepackage{graphicx}
\usepackage{color}

\newcommand{\text}[1]{\mbox{\scriptsize{#1}}}

\begin{document}

\title{The Role of Source Delocalization in the Development of Morphogen Gradients}
\author{Hamid Teimouri and Anatoly B. Kolomeisky}
\address{Department of Chemistry and Center for Theoretical Biological Physics, Rice University, Houston, Texas, 77005, USA}
\eads{\mailto{tolya@rice.edu}}

\begin{abstract}

Successful biological development via spatial regulation of cell differentiation relies on action of multiple signaling molecules that are known as morphogens. It is now well established that signaling molecules create non-uniform concentration profiles, morphogen gradients, that activate different genes, leading to patterning in the developing embryos.  The current view of the formation of morphogen gradients is that it is a result of complex reaction-diffusion processes that include the strongly localized production, diffusion and uniform degradation of signaling molecules. However, multiple experimental studies also suggest that the production of morphogen in many cases is delocalized. We develop a theoretical method that allows us to investigate the role of the delocalization in the formation of morphogen gradients. The approach is based on discrete-state stochastic models that can be solved exactly for arbitrary production lengths and production rates of morphogen molecules. Our analysis shows that the delocalization might have a strong effect on mechanisms of the morphogen gradient formation. The physical origin of this effect is discussed.

\end{abstract}


\newpage

\section{Introduction\label{sec1}}

One of the most important fundamental biological processes is a formation of a multi-cellular organism from a set of genetically identical cells. The central role in this process is played by  morphogens, which are signaling molecules that control the fate of biological cells \cite{book,lodish_book,wolpert_book,wolpert69,kerszberg98,lander07,crick70,kornberg12,rogers11}.  It is now widely accepted that these signaling molecules develop non-uniform concentration profiles, called morphogen gradients, that interact with embryo cells. The idea behind the morphogen gradients is that different concentrations of signaling molecules turn on different genes, leading to  complex patterning  observed in developed multi-cellular systems \cite{book,wolpert_book,wolpert69}.  A significant progress in understanding of  morphogen gradients  and how they function has been achieved in recent years when many quantitative investigations have appeared \cite{rogers11,tabata04,porcher10,zhou12,kicheva07,yu09,muller12,entchev00,drocco11,little11,spirov09,gregor07}. However, despite these advances, mechanisms of the formation of signaling molecules profiles remain not fully understood \cite{kornberg12,lander07}.

Several theoretical ideas were presented to explain the complex processes associated with the development of morphogen gradients  \cite{kornberg12,lander07}. The dominating view is based on a so-called synthesis-diffusion-degradation  (SDD) model \cite{kicheva07,porcher10,drocco12}. According to this approach, the production of signaling molecules starts at the specific localized region of the embryo, then signaling molecules diffuse along the cells. In addition, with equal probability they can be degraded after binding to specific receptors on the cells. As a result,  exponentially decaying concentration profiles are developed at long times.  Since qualitatively similar behavior is observed in many experimental systems, the SDD models have been widely utilized for understanding morphogen gradients \cite{porcher10,gregor07,kicheva07,yu09,drocco12,grimm10,entchev00,zhou12}. 

The majority of investigations of morphogen gradients formation that use the SDD models postulate that the signaling molecules  are produced from a sharply localized source \cite{berezhkovskii10,berezhkovskii11a,berezhkovskii11b,berezhkovskii11c,kolomeisky11,teimouri14,gordon13,muller13}. However, experimental observations suggest that in many biological systems the production region of the morphogens is delocalized \cite{porcher10}. Morphogens are protein molecules that are synthesized from the corresponding RNA molecules. So the production of these signaling molecules to a large degree is determined by the distribution of the corresponding RNA molecules. For one of the most intensely studied system, the formation of bicoid morphogen in early Drosophila embryo, it is known that the maternal RNA molecules are distributed over the region of size 30-50 $\mu$m,  while the total length of the embryo is of order of 400 $\mu$m \cite{porcher10}. Obviously, in this case the production area cannot be defined as sharply localized. This raises many queries on the role of the source production in the developing morphogen gradients. Specific  questions include: why the synthesis of signaling molecules is delocalized, why it is not produced over the whole embryo, and how the morphogen gradient depends on the spatial distribution of the source and on the synthesis rate?  At the same time, although some of these issues were discussed, a comprehensive theoretical analysis of the delocalization of signaling molecules synthesis  is not available  \cite{grimm10,dalessi12,deng10,berezhkovskii10,berezhkovskii11a,berezhkovskii11b,berezhkovskii11c}.
 
In this paper we present a theoretical investigation on the role of source production in the development of morphogen gradients. A theoretical approach for analyzing the formation of signaling molecules concentration profiles with arbitrary delocalization length and production rates is developed. Our method is based on discrete-state stochastic models that can be explicitly solved for arbitrary sets of parameters. We investigate several possible cases of the formation of signaling molecules concentration profiles to analyze the role of the synthesis of the signaling molecules. It is shown that the production might have a strong impact on the development of morphogen gradients.

The paper is organized as follows. In section 2, a general discrete-state stochastic SDD model with an arbitrary source distribution is presented and analyzed. In section 3, three different distributions are explicitly analyzed as illustrative examples. In this section the stationary profiles and local accumulation times are calculated.  Finally, in section 4, the important biophysical implications of the presented results are discussed and  conclusions are given.

\section{Theoretical Method}

We start our analysis by considering a general discrete-state stochastic  SDD model in one dimension as illustrated in figure 1.  Our system is semi-infinite. It is assumed that the synthesis of the morphogen particles is taking place only in the interval consisting of $L$ sites: see figure 1. Inside the source region, signalling molecules are produced at any site $m$ ($0 \le m \le L$) with a corresponding rate $Q_{m}$, and the total production rate is equal to $Q=\sum_{m=0}^{L} Q_{m}$. From any site morphogens can jump to nearest neighbors left or right sites with a rate $u$. The particles might be degraded at any site with a rate $k$. One can define a function $P(n,t;m)$ as the probability to find the morphogen at the site $n$ at time $t$ if the particle can only be produced at the site $m$ ($0 \le m \le L$). The temporal evolution of this probability is governed by the following master equations: 
\begin{eqnarray}\label{master2}
\frac{dP(m,t;m)}{dt}=Q_{m} \delta_{m,n}+u[P(m-1,t;m)+P(m+1,t;m)]\nonumber \\
-(2u+k)P(m,t;m)
\end{eqnarray}
for $n>0$, and
\begin{eqnarray}\label{master3}
\frac{dP(0,t;m)}{dt}=Q_{0} \delta_{m,1}+ uP(1,t;m)-(u+k)P(0,t;m)
\end{eqnarray}
for $n=0$. Here we use the fact that $\delta_{m,n}=1$ for $m=n$, and it is zero otherwise.

At long times we have $\frac{dP(n,t;m)}{dt}=0$, and these equations can be solved analytically, yielding the following stationary probability functions,  
\begin{eqnarray}\label{steady_profile1}
P^{(s)}_{<}(n;m)=\frac{Q_{m}[(k+\sqrt{k^2+4uk})x^{m-n}+(-k+\sqrt{k^2+4uk})x^{n+m})]}{(k+\sqrt{k^2+4uk})\sqrt{k^2+4uk}}\nonumber \\
\end{eqnarray}
for $ 0 \le n\le m$, and
\begin{eqnarray}\label{steady_profile2}
P^{(s)}_{>}(n;m)=\frac{Q_{m}[(k+\sqrt{k^2+4uk})x^{n-m}+(-k+\sqrt{k^2+4uk})x^{n+m})]}{(k+\sqrt{k^2+4uk})\sqrt{k^2+4uk}}\nonumber \\
\end{eqnarray}
for $ m\le n$.
In these expressions, the subscript $<$ $(>)$ corresponds to the case of $n \le m$ $(n>m)$, and a parameter $x$ ($0 < x < 1$) is defined as  
\begin{equation}\label{lambda}
x=(2u+k - \sqrt{k^{2} + 4uk})/(2u).
\end{equation}

\begin{figure}[ht]
\begin{center}
\includegraphics[width=0.95\linewidth]{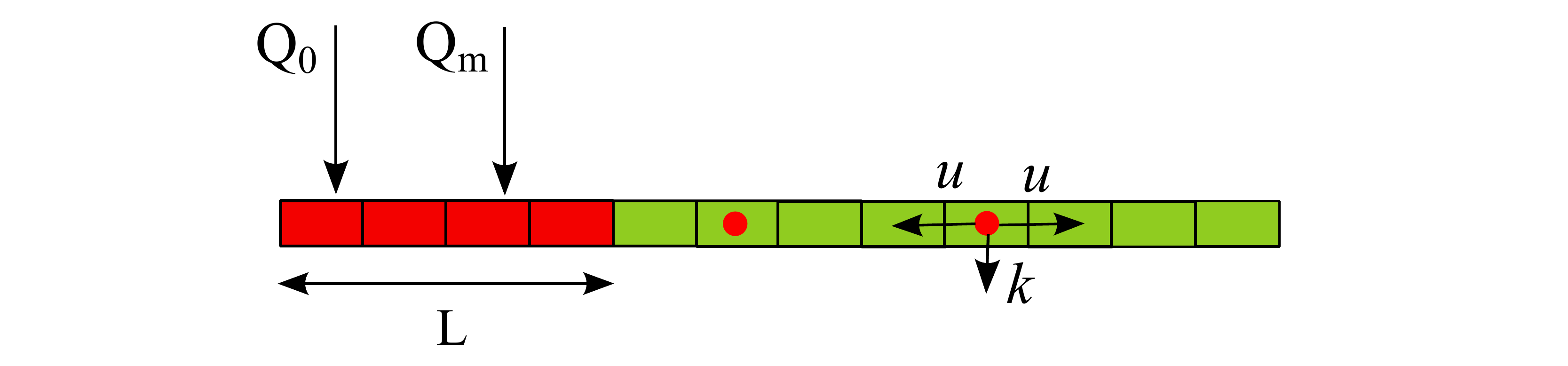}
\end{center}
\caption{ A schematic view of the one-dimensional discrete-state SDD model for the formation of the morphogen gradients. The production of morphogens is distributed over an interval of length $L$.  Signalling molecule are produced at the sites $1 \le m \le L$ (shown in red) with a rate $Q_{m}$. Particles can also diffuse along the lattice to the neighboring sites with a rate $u$, or they might be degraded with a rate $k$.}  
\end{figure}

This approach is very useful since it allows us to obtain the total concentration profiles from all producing sites using a kind of a superposition principle. In other words, the total probability $P(n,t)$ of finding the particle at site $n \ge 0$ at time $t$ can be expressed as a sum of the probabilities $P(n,t;m)$ with production at the specific site $m$ ($0 \le m \le L$). This is because the synthesis processes at each site are independent of each other. The general equations for concentration profiles are given by

\begin{equation}\label{super}
   P(n,t)= \left\{
     \begin{array}{ll}
      \sum\limits_{m=0}^{n}  P_{>}(n,t;m)+ \sum\limits_{m=n+1}^{L}  P_{<}(n,t;m), &  \mbox{for } 0\le n\le L;\\
      \sum\limits_{m=0}^{L}  P_{>}(n,t;m), &  \mbox{for } L\le n.
     \end{array}
   \right.
\end{equation} 
 
The explicit expressions can be easily obtained by employing equations (\ref{steady_profile1}) and (\ref{steady_profile2}).

One of the main goals of morphogen gradients is to transfer the  information. Most probably, it can be done well if the system is close to the stationary conditions. Then the important characteristics of morphogen gradients are times needed to achieve the steady state at specific spatial locations. These times are known as local accumulation times (LAT), and a theoretical framework for computing these quantities has been developed recently   \cite{berezhkovskii10}.   It can be done utilizing  local relaxation functions which are defined as
\begin{equation}\label{lat1}
R(n,t)=\frac{P(n,t)-P^{(s)}(n)}{P(n,t=0)-P^{(s)}(n)} =1-\frac{P(n,t)}{P^{(s)}(n)}.
\end{equation}
The physical meaning of these functions is that they represent the relative distance to the stationary state: at $t=0$ the distance is one, while at steady state  it is equal to zero. The explicit formulas for the local accumulation time can be derived then via  Laplace transformations of the local relaxation function, $\tilde{R}(n,s)=\int_{0}^{\infty}{R(n,t)e^{-st}dt}$ \cite{berezhkovskii10}, 
\begin{equation}\label{lat2}
t(n)=-\int_{0}^{\infty}{t\frac{\partial{R(n,t)}}{\partial{t}}e^{-st}dt} =\tilde{R}(n,s=0).
\end{equation}

\section{Illustrative Examples}

Our approach allows us to analyze the formation of morphogen gradients for arbitrary length of the production region and for arbitrary production rates. To explain it better, we illustrate the method by doing explicit calculations for three different production scenarios, which might be relevant for the morphogen gradients formation in real cellular conditions.

\subsection{Single Localized Source}

As a first example, we start with the case when the source of signaling molecules is localized at the site $m=m^{\prime}$,
\begin{equation}\label{delta}
Q_{m}=Q\delta(m-m^{\prime}).
\end{equation}
From equations (\ref{steady_profile1}) and  (\ref{steady_profile2}) we directly obtain 
\begin{equation}\label{localized_profile1}
\hspace{-20mm} P^{(s)}_{<}(n;m^{\prime})=\frac{Qx^{m^{\prime}}}{k+\sqrt{k^2+4uk}}\left[\frac{(k+\sqrt{k^2+4uk})x^{-n}+(-k+\sqrt{k^2+4uk})x^{n})}{\sqrt{k^2+4uk}}\right], 
\end{equation}
\begin{equation}\label{localized_profile2}
\hspace{-20mm} P^{(s)}_{>}(n;m^{'})=\frac{Qx^{n}}{k+\sqrt{k^2+4uk}}\left[\frac{(k+\sqrt{k^2+4uk})x^{-m^{'}}+(-k+\sqrt{k^2+4uk})x^{m^{'}}}{\sqrt{k^2+4uk}}\right].
\end{equation}
For $m^{\prime}=0$, as expected,  we recover the results obtained earlier \cite{kolomeisky11},
\begin{equation}\label{localized_profile3}
\hspace{-20mm} P^{(s)}(n;0)=\frac{2Qx^{n}}{k+\sqrt{k^2+4uk}}.
\end{equation}
These expressions indicate that the signaling molecules profiles are  exponentially decaying functions of the distance from the source for $n > m^{\prime}$, and the decay length, $\lambda=-1/\ln{x}$, is independent of the production rate $Q$. The resulting morphogen gradients are presented in figures 2 and 3 for the case of $m^{\prime}=0$ with various diffusion and degradation rates.

From relations (\ref{lat1}) and (\ref{lat2}) the explicit expressions for the local accumulation times can be evaluated. It is found that LAT are given by
\begin{eqnarray}\label{localized_time1}
\hspace{-20mm} t_{<}(n;m^{\prime})=\frac{1}{\sqrt{k^2+4uk}}\left[m^{\prime}+\frac{2u+k+\sqrt{k^2+4uk}}{k+\sqrt{k^2+4uk}}\right]\nonumber \\
+ \frac{n}{\sqrt{k^2+4uk}} \left[\frac{x^{n}(-k+\sqrt{k^2+4uk}) - x^{-n}(k+\sqrt{k^2+4uk})}{x^{-n}(k+\sqrt{k^2+4uk}) + x^{n}(-k+\sqrt{k^2+4uk})}\right]\nonumber \\
+\frac{2uk}{k^2+4uk}\left[\frac{(x^{n}-x^{-n})}{x^{-n}(k+\sqrt{k^2+4uk}) + x^{n}(-k+\sqrt{k^2+4uk})}\right],
\end{eqnarray}
\begin{eqnarray}\label{localized_time2}
\hspace{-20mm} t_{>}(n;m^{\prime})=\frac{1}{\sqrt{k^2+4uk}}\left[n+\frac{2u+k+\sqrt{k^2+4uk}}{k+\sqrt{k^2+4uk}}\right]\nonumber \\
+ \frac{m^{\prime}}{\sqrt{k^2+4uk}} \left[\frac{x^{m^{\prime}}(-k+\sqrt{k^2+4uk}) - x^{-m^{\prime}}(k+\sqrt{k^2+4uk})}{x^{-m^{\prime}}(k+\sqrt{k^2+4uk}) + x^{m^{\prime}}(-k+\sqrt{k^2+4uk})}\right]\nonumber \\
+\frac{2uk}{k^2+4uk}\left[\frac{(x^{m^{\prime}}-x^{-m^{\prime}})}{x^{-m^{\prime}}(k+\sqrt{k^2+4uk}) + x^{m^{\prime}}(-k+\sqrt{k^2+4uk})}\right].
\end{eqnarray}
Again, for the source localized at the origin, $m^{\prime}=0$, our results reproduce the already known expression \cite{kolomeisky11},
\begin{eqnarray}\label{localized_time0}
\hspace{-20mm} t(n)=\frac{1}{\sqrt{k^2+4uk}}\left[n+\frac{2u+k+\sqrt{k^2+4uk}}{k+\sqrt{k^2+4uk}}\right].
\end{eqnarray}
LAT for the formation of morphogen gradients with $m^{\prime}=0$ for various diffusion and degradation rates are presented in figures 4 and 5. One can see that the local accumulation times for the sharply localized source linearly grow with the distance from the source. This can be explained by invoking the idea that the degradation acts as an effective potential that pushes the signaling molecules away from the source \cite{kolomeisky11}, leading to effectively driven diffusion of morphogens in the system.

\begin{figure}[ht]
\begin{center}
\includegraphics[width=0.6\linewidth, angle =-90 ]{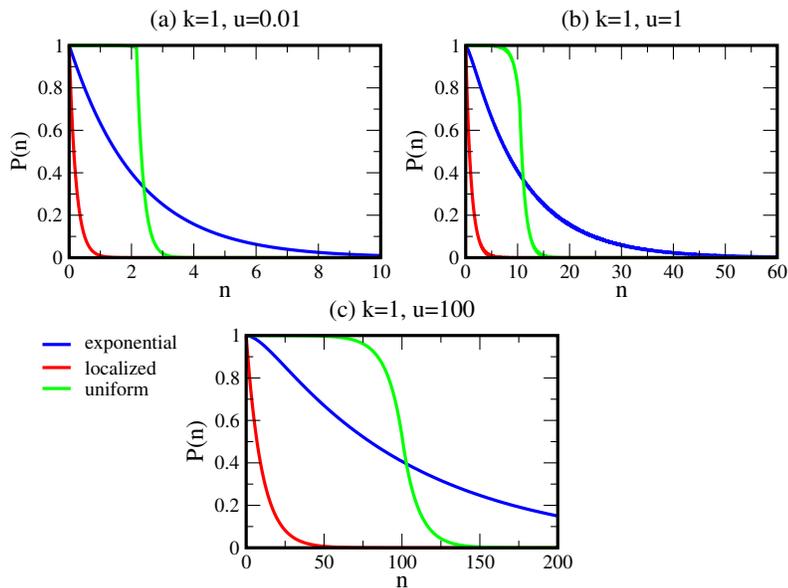}
\end{center}
\caption{Steady-state density profiles as a function of the distance from the origin.  Red curves correspond to the single localized source at $m^{\prime}=0$. Green curves correspond to uniform production rates  along the finite interval. Blue curves correspond to exponentially decaying production rates along the semi-infinite interval. $L=\lambda_{s}=10\lambda$ is assumed for the single localized and uniform productions, while $\lambda_{s}=10\lambda$ and $L \rightarrow \infty$ are  assumed for the exponentially distributed productions.  (a) Fast degradation rates with $k=1$, $u=0.01$; (b) Comparable diffusion and degradation rates with $k=u=1$; and (c) Fast diffusion rates with $k=1$, $u=100$. }  
\end{figure}

\begin{figure}[ht]
\begin{center}
\includegraphics[width=0.6\linewidth, angle =-90 ]{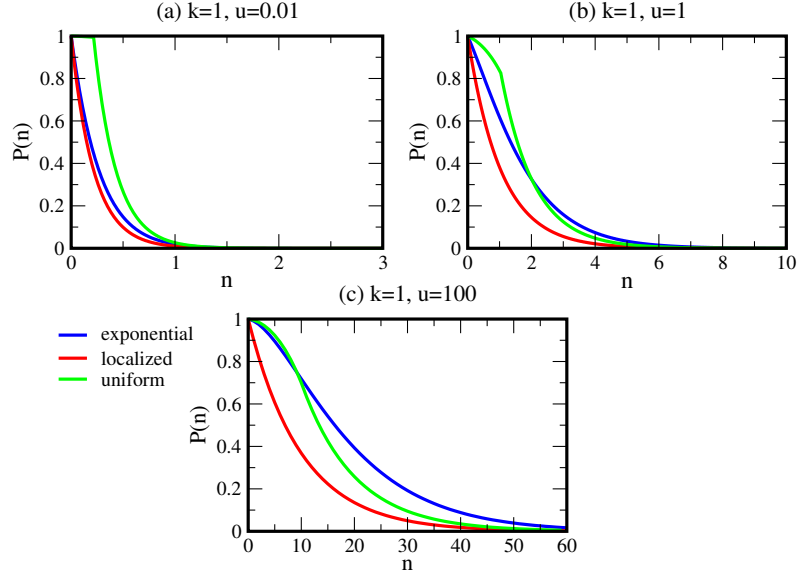}
\end{center}
\caption {Steady-state density profiles as a function of the distance from the origin. Red curves correspond to the single localized source at $m^{\prime}=0$. Green curves correspond to uniform production rates  along the finite interval. Blue curves correspond to exponentially decaying production rates along the semi-infinite interval. $L=\lambda_{s}=10\lambda$ is assumed for the single localized and uniform productions, while $\lambda_{s}=\lambda$ and $L \rightarrow \infty$ are  assumed for the exponentially distributed productions. (a) Fast degradation rates with $k=1$, $u=0.01$; (b) Comparable diffusion and degradation rates with $k=u=1$; and (c) Fast diffusion rates with $k=1$, $u=100$. }  
\end{figure}

\subsection{Uniformly Distributed Production Over the Finite Interval}

In another example, we consider  uniformly distributed production of signaling molecules  along a finite interval of length $L$. This distribution can be represented as follows,
\begin{eqnarray}
    Q_{m}= \left\{
     \begin{array}{ll}
      \frac{Q}{L+1},    \quad 0\le m\le L; \\
      0,                \quad        m >L .
     \end{array}
   \right.
\end{eqnarray} 
In this case, the synthesis rates are the same for all production sites. Applying equation (\ref{super}), one can  obtain explicit expressions for the stationary profiles. They are different depending on the position of the lattice site with respect to the production region.  Inside the production area we have
{\footnotesize 
\begin{eqnarray}\label{interval_profile1}
P^{(s)}_{<}(n)=\frac{Qx^{n}}{(L+1)(k+\sqrt{k^2+4uk})}\nonumber \\
\times \left [\frac{(k+\sqrt{k^2+4uk})(x-x^{-n}-x^{1-n}+x^{L-2n+1})+(-k+\sqrt{k^2+4uk})(-1+x^{L+1})}{(x-1)\sqrt{k^2+4uk}}\right], \nonumber \\
\end{eqnarray}}
while outside it can be shown that
{\footnotesize 
\begin{eqnarray}\label{interval_profile2}
P^{(s)}_{>}(n)=\frac{Qx^{n}}{(L+1)(k+\sqrt{k^2+4uk})}\nonumber \\
\times \left[\frac{(k+\sqrt{k^2+4uk})(x-x^{-L})+(-k+\sqrt{k^2+4uk})(-1+x^{L+1})}{(x-1)\sqrt{k^2+4uk}}\right]. 
\end{eqnarray}}
The maximal value of the concentration  is achieved for $n=0$,
\begin{equation}\label{interval_profile1_ampli}
P^{(s)}_{<}(n=0)=\frac{2 Q(-1+x^{L+1})}{(x-1)(k+\sqrt{k^2+4uk})(L+1)}.
\end{equation}
The concentration profiles normalized by this maximal value are presented in figures 2 and 3.

A stationary-state behavior of the morphogen gradients with the uniform production along the finite interval can be generally described in the following way. The concentration of signaling molecules is large and almost constant in the production region, and then it exponentially decays outside of the synthesis area. The transition between these two regimes depends on parameters of the system. Decreasing the diffusion and/or increasing the degradation rate makes this crossover sharper. In addition, the transition is more abrupt when the production length $L$ is much larger than the decay length $\lambda$ (see figure 2 where $L=10 \lambda$ is utilized), while for comparable  $L$ and $\lambda$ (see figure 3 where $L=\lambda$ is utilized) the concentration profile decays more smoothly.  

It is interesting to analyze the behavior of the  profiles in the  asymptotic limit of  very large production intervals, $L \rightarrow \infty$,
{\footnotesize 
\begin{eqnarray}\label{interval_profile_infty}
\hspace{-20mm}P^{(s)}(n,L\rightarrow \infty) \simeq \frac{Q(1+x)}{L(1-x)\sqrt{k^2+4uk}}.
\end{eqnarray}}
This corresponds to the case when the morphogens are synthesized all over the embryo  with the same production rates. As expected, the concentration profile becomes uniform, and this is the reason why there is  no dependence on the position $n$ in this equation. Obviously, the morphogen gradient cannot be established in such systems, and this is not a realistic situation for biological development. This might be the reason why the production region does not occupy the whole embryo.  

As was discussed above, the local accumulation times for uniform production along the finite intervals can be found from equations (\ref{lat1}) and (\ref{lat2}). LAT for sites inside the production region are given by
{\footnotesize 
\begin{eqnarray}\label{interval_time1}
\hspace{-20mm}t_{<}(n)&=&\frac{1}{\sqrt{k^2+4uk}}\left[n+\frac{2u+k+\sqrt{k^2+4uk}}{k+\sqrt{k^2+4uk}}\right]\nonumber \\
\hspace{-35mm}&+& \frac{1}{\sqrt{k^2+4uk}}\left[\frac{(nx^{-n}-(1-n)x^{1-n} + (L-2n+1)x^{L-2n+1})(k+\sqrt{k^2+ 4uk})+ Lx^{L+1}(-k+\sqrt{k^2+4uk})}{(x-x^{-n}-x^{1-n} + x^{L-2n+1} )(k+\sqrt{k^2+4uk}) - (1-x^{L+1})(-k+\sqrt{k^2+4uk})}\right]\nonumber \\
\hspace{-35mm}&+& \frac{2uk}{k^2+4uk}\left[\frac{-x+ x^{-n} + x^{1-n} - x^{L-2n+1} -1 + x^{L+1}}{(x-x^{-n}-x^{1-n} + x^{L-2n+1} )(k+\sqrt{k^2+4uk}) - (1-x^{L+1})(-k+\sqrt{k^2+4uk})}\right]\nonumber \\
\hspace{-35mm}&+& \frac{x}{(x-1)\sqrt{k^2+4uk}}\left[\frac{(1-x^{L})(-k+\sqrt{k^2+4uk}) + (x^{-n} + x^{1-n} - x^{L-2n+1} -1)(k+\sqrt{k^2+4uk})}{(x-x^{-n}-x^{1-n} + x^{L-2n+1} )(k+\sqrt{k^2+4uk}) - (1-x^{L+1})(-k+\sqrt{k^2+4uk})}\right];
\end{eqnarray}}
while for the outside region we have
{\footnotesize 
\begin{eqnarray}\label{interval_time2}
\hspace{-8mm}t_{<}(n)&=&\frac{1}{\sqrt{k^2 + 4uk}}\left[n+\frac{2u+k+\sqrt{k^2+4uk}}{k+\sqrt{k^2+4uk}}\right]\nonumber \\
\hspace{-35mm}&+& \frac{L}{\sqrt{k^2 + 4uk}}\left[\frac{x^{-L}(k+\sqrt{k^2+4uk})+ x^{L+1}(-k+\sqrt{k^2+4uk})}{(x- x^{-L} )(k+\sqrt{k^2 + 4uk}) - (1-x^{L+1})(-k+\sqrt{k^2  + 4uk})}\right]\nonumber \\
\hspace{-35mm}&+& \frac{2uk}{k^2+4uk}\left[\frac{x^{-L}-x-1+ x^{L+1}}{(x- x^{-L} )(k+\sqrt{k^2 + 4uk}) - (1-x^{L+1})(-k+\sqrt{k^2 + 4uk})}\right]\nonumber \\
\hspace{-35mm}&+& \frac{x}{(x-1)\sqrt{k^2 + 4uk}}\left[\frac{(x^{-L}-1)(k+\sqrt{k^2 + 4uk})+ (1-x^{L})(-k+\sqrt{k^2 + 4uk})}{(x- x^{-L} )(k+\sqrt{k^2+4uk}) - (1-x^{L+1})(-k+\sqrt{k^2+4uk})}\right].\nonumber \\
\end{eqnarray}}

The results for local accumulation times are presented in figures 4 and 5. For the uniform production along the finite interval the LAT usually consist of two parts. Inside the production area LAT is almost constant as a function of the distance from the origin. At the same time, outside of the production area LAT is linearly growing. Again, arguments using the role of the degradation as an effective potential can be employed \cite{kolomeisky11}. Outside of the synthesis region, the degradation creates the potential that moves particles away from the production area. This effective potential leads to strong biased diffusion and consequently linearly growing local accumulation times. This behavior is very similar to the case of the single localized source discussed above.  Inside the synthesis region the dynamics is different. The degradation is compensated by synthesis and diffusion from neighboring cells, leading to mostly uniform accumulation times.

\begin{figure}[ht]
\begin{center}
\includegraphics[width=0.6\linewidth, angle =-90 ]{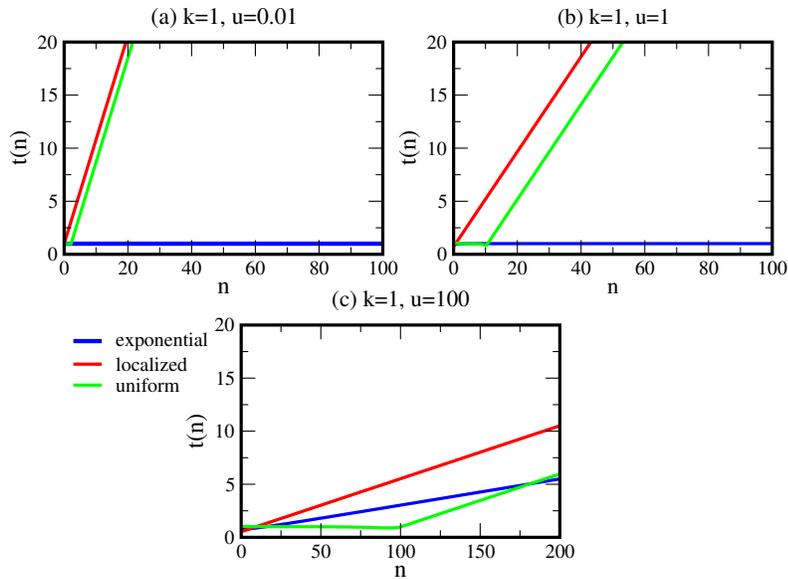}
\caption{Local accumulation times as a function of the distance from the origin. Red curves correspond to the single localized source at $m^{\prime}=0$. Green curves correspond to uniform production rates  along the finite interval. Blue curves correspond to exponentially decaying production rates along the semi-infinite interval. $L=\lambda_{s}=10\lambda$ is assumed for the single localized and uniform productions, while $\lambda_{s}=10\lambda$ and $L \rightarrow \infty$ are  assumed for the exponentially distributed productions. (a) Fast degradation rates with $k=1$, $u=0.01$; (b) Comparable diffusion and degradation rates with $k=u=1$; and (c) Fast diffusion rates with $k=1$, $u=100$.}  
\end{center}
\end{figure}

\begin{figure}[ht]
\begin{center}
\includegraphics[width=0.65\linewidth, angle =-90 ]{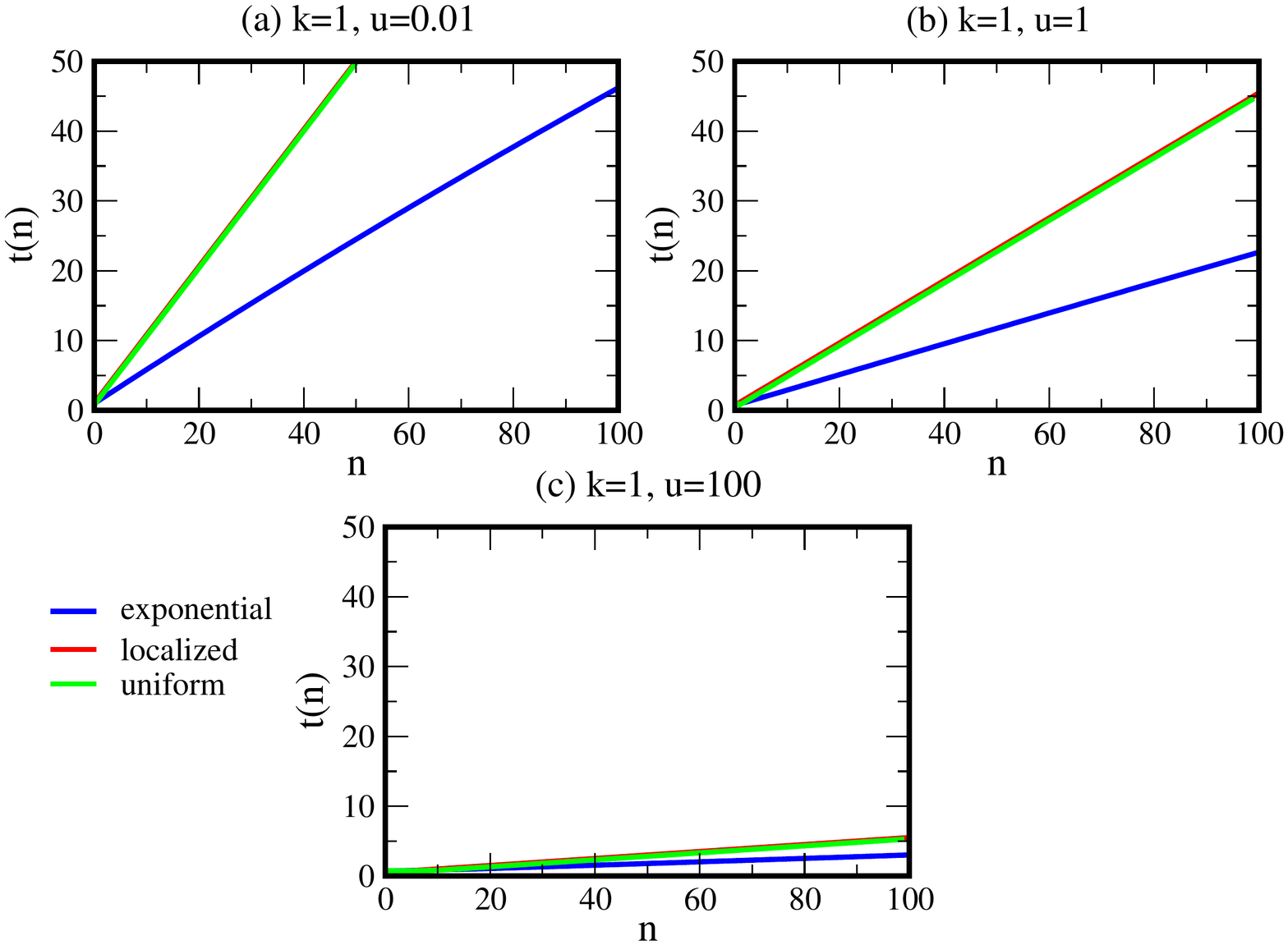}
\end{center}
\caption{Local accumulation times as a function of the distance from the origin. Red curves correspond to the single localized source at $m^{\prime}=0$. Green curves correspond to uniform production rates  along the finite interval. Blue curves correspond to exponentially decaying production rates along the semi-infinite interval. $L=\lambda_{s}=\lambda$ is assumed for the single localized and uniform productions, while $\lambda_{s}=\lambda$ and $L \rightarrow \infty$ are  assumed for the exponentially distributed productions. (a) Fast degradation rates with $k=1$, $u=0.01$; (b) Comparable diffusion and degradation rates with $k=u=1$; and (c) Fast diffusion rates with $k=1$, $u=100$.}  
\end{figure}

\subsection{Exponentially Distributed Production along the  Semi-Infinite Interval}

In the final example, we consider the exponentially distributed production along the interval of length $L$. In this case, the synthesis rates can be written as
\begin{eqnarray}\label{exponential_dist}
Q_{m}=\frac{Q(1-z)z^{m}}{1-z^{L+1}},
\end{eqnarray}
where $0 \le m \le L$ and a parameter $z$  ($0 < z <1$) is introduced to characterize the exponential distribution of synthesis rates. One can see that the relation $\sum_{m=0}^{L} Q_{m}=Q$ is satisfied.  We can also define a decay length $\lambda_{s}$ for this exponential distribution of the production rates,
\begin{equation}
\lambda_{s}=-1/\ln{z}.
\end{equation}

Detailed calculations of the stationary profiles for exponentially distributed production rates and for arbitrary length of the source region are given in the Appendix. Here we report the final results for the case of the semi-infinite source interval, i.e., in the limit of $L \rightarrow \infty$,
{\footnotesize 
\begin{eqnarray}\label{exponential_profile_infty}
\hspace{-15mm}& P^{(s)}(n)=\frac{Q(1-z)}{\sqrt{k^2+4uk}}\left[\frac{(x^{n+1}-zx^{n+2}-z^{n+1}+x^{2}z^{n+1})}{(x-z)(1-xz)}\right ]  \nonumber \\
\hspace{-15mm}& + \frac{Q(1-z)}{\sqrt{k^2+4uk}}\left[\frac{(-k+\sqrt{k^2+4uk})x^{n}}{(k+\sqrt{k^2+4uk})(1-xz)}\right].
\end{eqnarray}
}
The maximal value of the concentration profile is achieved again for $n=0$,
\begin{equation}\label{interval_exp_ampli}
P^{(s)}(n=0)=\frac{2Q(1-z)}{(1-xz)(k+\sqrt{k^2+4uk})}.
\end{equation}

It is interesting to consider a special limiting case of this problem when the diffusion along the lattice cells is much faster than the degradation rate, $u \gg k$, while the decay length $\lambda_{s}$ for the exponential production rates is also large. This corresponds to a continuum limit of our problem, and it was already discussed in the literature \cite{berezhkovskii10,berezhkovskii11a,berezhkovskii11b,berezhkovskii11c}. In this limit, we have the following asymptotic relations for important parameters $x$ and $z$,
\begin{equation}\label{x_expansion}
x \simeq 1- \frac{1}{\lambda}, 
\end{equation}
\begin{equation}\label{z_expansion}
  z \simeq 1- \frac{1}{\lambda_{s}}.
\end{equation}
Using these results, one can easily show that the equation (\ref{exponential_profile_infty}) reduces to the following form,
\begin{eqnarray}\label{exponential_continuum}
P(n)=\frac{Q\lambda^{2}}{u(\lambda^{2}-\lambda_{s}^{2})}( \lambda e^{-n/\lambda}-\lambda_{s} e^{-n/\lambda_{s}}).
\end{eqnarray}
This is exactly the concentration profile obtained previously in the continuum SDD model with the exponentially decaying distribution of the production rates \cite{berezhkovskii11c}.

The normalized concentration profiles for exponentially decaying production along the semi-indite interval ($L \gg 1$)  are plotted in figures 2 and 3. For all sets of the parameters these profiles are monotonically decreasing functions of the distance from the origin, although the decay is slower for larger diffusion rates and/or for smaller degradation rates. This allows to increase significantly the range of morphogen action, as compared to systems with the sharply localized source, while still keeping the concentration profile of signaling molecules to be non-uniform. As expected, accelerating the exponential decay of production rates (lowering of $\lambda_{s}$) diminishes this effect as one can see by comparing figures 2 and 3.  As expected, for small $\lambda_{s}$ the concentration profiles of morphogen gradients with  exponentially distributed productions become similar to the morphogen gradients with the single localized source.

The calculations for local accumulation times for exponentially decaying synthesis of signalling molecules  and for arbitrary production lengths $L$  are also given in the Appendix. We are interested in the large $L$ values that correspond to the semi-infinite interval. In this case, we obtain the following expression for LAT,
{\footnotesize
\begin{eqnarray}\label{exponential_time_infty}
\hspace{-25mm}t(n) = \frac{n}{\sqrt{k^{2}+4uk}}\left[\frac{(-k+\sqrt{k^2+4uk})(x^{n+1}-z x^{n})+(k+\sqrt{k^2+4uk})(x^{n+1}-zx^{n+2})}{(k + \sqrt{k^{2} +4uk})( x^{n+1}-zx^{n+2}- z^{n+1} + x^{2}z^{n+1}) + (-k+\sqrt{k^{2}+4uk})(x^{n+1}-z x^{n})
}\right] \nonumber\\
\hspace{-30mm} +\frac{2uk}{k^{2}+4uk}\left[\frac{ z^{n+1} - z x^{n} + z x^{n+2} -x^{2} z^{n+1} }{(k + \sqrt{k^{2} +4uk})( x^{n+1}-zx^{n+2}- z^{n+1} + x^{2}z^{n+1}) + (-k+\sqrt{k^{2}+4uk})(x^{n+1}-z x^{n})
}\right] \nonumber\\
\hspace{-30mm} + \frac{1}{\sqrt{k^{2}+4uk}}\left[\frac{(k+\sqrt{k^2+4uk})(x^{n+1} - 2zx^{n+2} + 2x^{2}z^{n+1} )+(-k+\sqrt{k^2+4uk})x^{n+1} }{(k + \sqrt{k^{2} +4uk})( x^{n+1}-zx^{n+2}- z^{n+1} + x^{2}z^{n+1}) + (-k+\sqrt{k^{2}+4uk})(x^{n+1}-z x^{n})
}\right] \nonumber\\
+\frac{1}{\sqrt{k^{2}+4uk}}\left[\frac{2u+k+\sqrt{k^{2}+4uk}}{k+ \sqrt{k^{2}+4uk}}+ \frac{xz}{1-xz}-\frac{x}{x-z}\right].
\end{eqnarray}}

The results for LAT for the exponentially distributed production of the morphogens are illustrated in figures 4 and 5. In all cases the local accumulation times increase with the distance from the origin, as was found for other production scenarios of morphogen gradients considered in this work. The LAT growth is mostly linear since our arguments about the effective potential due to degradation still can be applied. But what is different from other production mechanisms is the fact that these times are usually much smaller, especially at large distances from the origin (larger than $L$ or $\lambda_{s}$). This can be understood if we recall that main contribution to the local accumulation times at given spatial position  is due to the first-passage  (arrival) times to this location starting from the source region \cite{kolomeisky11}. In the exponentially distributed production of signaling molecules along the semi-infinite interval  the source is everywhere,  so that the contribution of the arrival times is mostly negligible. This significantly lowers the time to reach the stationary-state concentration at the given spatial location.

\section{Summary and Conclusions}

We developed a theoretical framework for investigating the role of synthesis of signaling molecules in the formation of concentration profiles that are critically important in the biological development. Our analysis is based on discrete-state stochastic models for complex reaction-diffusion processes associated with the formation of morphogen gradients. It allows us to test the effect of the production in systems with arbitrary source lengths and synthesis rates by calculating stationary profiles and the times to achieve the stationary concentrations at specific locations.  By analyzing several different systems we found that the spatial distributions of the sources and the production speeds  have a strong effect in the development of morphogen gradients.

To understand the role of the production one might recall that there are two main requirements for successful function of the morphogen gradients in the biological systems \cite{lander07}. The first one is to deliver the information to as many as possible  embryo cells  about their future fates. This can be done if observable concentrations of signaling molecules can be found as far as possible from the source. The second function is to ensure that different genes can be controllably turned on  in the neighboring embryo cells. This can be accomplished by producing sharp boundaries in the concentration profiles of signaling molecules at specific locations. Our analysis suggests that the morphogen gradients produced with the single localized sources generally are not able to satisfy these requirements.  At the same time, the morphogen gradients developed from the delocalized sources with various synthesis rates are capable to do all these tasks successfully.  

Furthermore, the presented theoretical method  provides  a simple physical-chemical explanation on the role of delocalizations in the formation of signaling molecules concentration profiles. The delocalization effectively leads to faster diffusion along the productions regions, and it also shortens the arrival times to specific locations. As a result, the range of morphogen gradients with delocalized sources increases, while the times to reach the stationary states at specific locations become smaller. All of these properties make the morphogen gradients more efficient and robust. At the same, increasing the production area to the whole embryo is not reasonable since it will be difficult to sustain the non-uniform concentration profiles.   

Although our theoretical approach gives a clear physical picture on the effect of production in the development of morphogen gradients, one should note that our model is oversimplified.  It neglects many  crucial factors and processes  in the biological development  such as  a three-dimensional nature of the embryo, inhomegeneity and complex cooperative mechanisms in the degradation, as well as variable mechanical responses of the embryo cells. It will be critically important to test the proposed theoretical picture in experimental studies and in more advanced theories.

\section*{Acknowledgments}

The work was supported by grants from the Welch Foundation (C-1559), from the NSF (Grant CHE-1360979) and by the Center for Theoretical Biological Physics sponsored by the NSF (Grant PHY-1427654).

\section*{Appendix}

In this Appendix we present detailed calculations of the stationary profiles and LATs for the exponential distribution of production rates. Using a superposition principle, the total stationary profile can be written as
{\footnotesize 
\begin{eqnarray}\label{appned1}
\hspace{-30mm}P^{(s)}_{<}(n)=\frac{x^{n}}{\sqrt{k^2+4uk}(k+\sqrt{k^2+4uk})}\left[(k+\sqrt{k^2+4uk})\sum_{m=0}^{n}Q_{m}x^{-m} + (-k+\sqrt{k^2+4uk})\sum_{m=0}^{n}Q_{m}x^{m}\right]\nonumber\\ 
\hspace{-15mm} + \frac{\left[(k+\sqrt{k^2+4uk})x^{-n}+(-k+\sqrt{k^2+4uk})x^{n}\right]}{\sqrt{k^2+4uk}(k+\sqrt{k^2+4uk})}\left[\sum_{m=n+1}^{L} Q_{m}x^{m}\right],
\end{eqnarray}}
for the sites inside the production area. For the sites outside of the production  we have
{\footnotesize 
\begin{eqnarray}\label{appned2}
\hspace{-30mm}P^{(s)}_{>}(n)=\frac{x^{n}}{\sqrt{k^2+4uk}(k+\sqrt{k^2+4uk})}\left[(k+\sqrt{k^2+4uk})\sum_{m=0}^{L}Q_{m}x^{-m} + (-k+\sqrt{k^2+4uk})\sum_{m=0}^{L}Q_{m}x^{m}\right].\nonumber\\ 
\end{eqnarray}}
The summations over $m$ for different intervals can be performed in the following way,
\begin{eqnarray}\label{summation1}
\sum^{L}_{m=0} Q_{m}x^{m} =\frac{Q(1-z)}{1-z^{L+1}}\left [\frac{1-(xz)^{L+1}}{1-xz} \right];
\end{eqnarray}
\begin{eqnarray}\label{summation2}
\sum^{L}_{m=0} Q_{m}x^{-m} =\frac{Q(1-z)}{1-z^{L+1}}\left [\frac{x-z(z/x)^{L}}{x-z} \right];
\end{eqnarray}
\begin{eqnarray}\label{summation3}
\sum^{n}_{m=0} Q_{m}x^{-m} = \frac{Q(1-z)}{1-z^{L+1}}\left [\frac{x-z(z/x)^{n}}{x-z} \right];
\end{eqnarray}
\begin{eqnarray}\label{summation4}
\sum^{L}_{m=n+1} Q_{m}x^{m} = \frac{Q(1-z)}{1-z^{L+1}}\left [\frac{(xz)^{n+1}-(xz)^{L+1}}{1-xz} \right].
\end{eqnarray}
Substituting these expressions into Eqs. (\ref{appned1}) and (\ref{appned2}), we obtain for the sites inside the source region,
{\footnotesize 
\begin{eqnarray}\label{exponential_profile1}
\hspace{-35mm}& P^{s}_{<}(n)=\frac{Q(1-z)}{\sqrt{k^2+4uk}(1-z^{L+1})}\left [\frac{(x^{n+1}-zx^{n+2}-z^{L+1}x^{L+2-n}+z^{L+2}x^{L+1-n}-z^{n+1}+x^{2}z^{n+1})}{(x-z)(1-xz)}\right] \nonumber \\
\hspace{-15mm}& + \frac{Q(1-z)}{\sqrt{k^2+4uk}(1-z^{L+1})}\left[ \frac{(-k+\sqrt{k^2+4uk})(1-(xz)^{L+1})(x^{n+1}-z x^{n})}{(k+\sqrt{k^2+4uk})(x-z)(1-xz)}\right];
\end{eqnarray}
}
while for the outside region we have
{\footnotesize 
\begin{eqnarray}\label{exponential_profile2}
\hspace{-35mm}& P^{s}_{>}(n)=\frac{Q(1-z)}{\sqrt{k^2+4uk}(1-z^{L+1})}\left [\frac{(x^{n+1}-zx^{n+2}-z^{L+1}x^{-L+n}+z^{L+2}x^{-L+1+n})}{(x-z)(1-xz)}\right] \nonumber \\
\hspace{-15mm}& + \frac{Q(1-z)}{\sqrt{k^2+4uk}(1-z^{L+1})}\left[ \frac{(-k+\sqrt{k^2+4uk})(1-(xz)^{L+1})(x^{n+1}-z x^{n})}{(k+\sqrt{k^2+4uk})(x-z)(1-xz)}\right].
\end{eqnarray}
}

The corresponding expressions for LATs  for the source region can be written as
{\footnotesize
\begin{eqnarray}\label{exponential_time<}
\hspace{-15mm}t_{<}(n) = \frac{n}{\sqrt{k^{2}+4uk}}\left[\frac{(-k+\sqrt{k^2+4uk})(1-(xz)^{L+1})(x^{n+1}-z x^{n})+(k+\sqrt{k^2+4uk})\Psi_{<}(n)}{\Delta_{<}(n)}\right] \nonumber\\
\hspace{-30mm} +\frac{2uk}{k^{2}+4uk}\left[\frac{(1-(xz)^{L+1})( x^{n+1} - z x^{n}) -x^{n+1} + z x^{n+2} +   z^{L+1} x^{L-n+2}- z^{L+2} x^{L-n+1} + z^{n+1} -x^{2} z^{n+1} }{\Delta_{<}(n)}\right] \nonumber\\
\hspace{-30mm} + \frac{L}{\sqrt{k^{2}+4uk}}\left[\frac{(k+\sqrt{k^2+4uk})(z^{L+2}x^{L-n+1}-z^{L+1}x^{L-n+2})-(-k+\sqrt{k^2+4uk})(x^{n+1}-z x^{n})(xz)^{L+1} }{\Delta_{<}(n)}\right] \nonumber\\
\hspace{-30mm} + \frac{1}{\sqrt{k^{2}+4uk}}\left[\frac{(k+\sqrt{k^2+4uk})\Phi_{<}(n)+(-k+\sqrt{k^2+4uk})((x^{n+1}(1-(xz)^{L+1}) -(x^{n+1}-z x^{n})(xz)^{L+1})}{\Delta_{<}(n)}\right] \nonumber\\
+\frac{1}{\sqrt{k^{2}+4uk}}\left[\frac{2u+k+\sqrt{k^{2}+4uk}}{k+ \sqrt{k^{2}+4uk}} + \frac{xz}{1-xz}-\frac{x}{x-z}\right],
\end{eqnarray}}
where we defined new auxiliary functions
\begin{eqnarray}\label{Delta_<}
\hspace{-20mm}\Delta_{<}(n) =(k + \sqrt{k^{2} +4uk})( x^{n+1} - zx^{n+2} - z^{L+1} x^{L+2-n} + z^{L+2} x^{L+1-n} - z^{n+1} + x^{2}z^{n+1}) \nonumber \\
+ (-k+\sqrt{k^{2}+4uk})(1-(xz)^{L+1})(x^{n+1}-z x^{n});
\end{eqnarray}
and
\begin{eqnarray}\label{Psi1}
\hspace{-20mm}\Psi_{<}(n) = x^{n+1} - zx^{n+2} + z^{L+1} x^{L+2-n} - z^{L+2} x^{L+1-n}, 
\end{eqnarray}
\begin{eqnarray}\label{Phi1}
\hspace{-20mm}\Phi_{<}(n) = x^{n+1} - 2zx^{n+2} - 2z^{L+1} x^{L+2-n}+ 2x^{2}z^{n+1} + z^{L+2} x^{L+1-n}.
\end{eqnarray}

Similar calculations for LAT for the region outside of the source area produce
{\footnotesize
\begin{eqnarray}\label{exponential_time>}
\hspace{-15mm}t_{>}(n) = \frac{n}{\sqrt{k^{2}+4uk}}\left[\frac{(-k+\sqrt{k^2+4uk})(1-(xz)^{L+1})(x^{n+1}-z x^{n})+(k+\sqrt{k^2+4uk})\Psi_{>}(n)}{\Delta_{>}(n)}\right] \nonumber\\
\hspace{-30mm} +\frac{2uk}{k^{2}+4uk}\left[\frac{(1-(xz)^{L+1})( x^{n+1} - z x^{n}) -x^{n+1} + z x^{n+2} +   z^{L+1} x^{L-n+2}- z^{L+2} x^{L-n+1} + z^{n+1} -x^{2} z^{n+1} }{\Delta_{>}(n)}\right] \nonumber\\
\hspace{-30mm} + \frac{L}{\sqrt{k^{2}+4uk}}\left[\frac{(k+\sqrt{k^2+4uk})(z^{L+1}x^{L-n}-z^{L+2}x^{L-n+1})
-(-k+\sqrt{k^2+4uk})(x^{n+1}-z x^{n})(xz)^{L+1} }{\Delta_{>}(n)}\right] \nonumber\\
\hspace{-30mm} + \frac{1}{\sqrt{k^{2}+4uk}}\left[\frac{(k+\sqrt{k^2+4uk})\Phi_{>}(n)+(-k+\sqrt{k^2+4uk})((x^{n+1}(1-(xz)^{L+1}) -(x^{n+1}-z x^{n})(xz)^{L+1})}{\Delta_{>}(n)}\right] \nonumber\\
+\frac{1}{\sqrt{k^{2}+4uk}}\left[\frac{2u+k+\sqrt{k^{2}+4uk}}{k+ \sqrt{k^{2}+4uk}} 
+ \frac{xz}{1-xz}-\frac{x}{x-z}\right],
\end{eqnarray}}
where another set of auxiliary functions is introduced,
\begin{eqnarray}\label{Delta_>}
\hspace{-20mm}\Delta_{>}(n) =(k + \sqrt{k^{2} +4uk})((x^{n+1}-zx^{n+2}-z^{L+1}x^{-L+n}+z^{L+2}x^{-L+1+n}) \nonumber \\
+ (-k+\sqrt{k^2+4uk})(1-(xz)^{L+1})(x^{n+1}-z x^{n});
\end{eqnarray}
and
\begin{eqnarray}\label{Psi2}
\hspace{-20mm}\Psi_{>}(n) = x^{n+1} - zx^{n+2} - z^{L+1} x^{n-L} + z^{L+2} x^{n-L+1},
\end{eqnarray}
\begin{eqnarray}\label{Phi2}
\hspace{-20mm}\Phi_{>}(n) = x^{n+1} - 2zx^{n+2} + z^{L+2} x^{n-L+1}.
\end{eqnarray}

Taking the limit of $L \rightarrow \infty$ in equations (\ref{exponential_profile1}) and (\ref{exponential_time<}) we obtain the expressions for the stationary-state profiles and the local accumulation times used in the main text of the paper.

\end{document}